\documentclass[aip,preprint,
amsmath,amssymb]{revtex4-1}

\usepackage{graphicx}
\usepackage{dcolumn}
\usepackage{bm}

\begin{document}

\title{Study of loss in superconducting coplanar waveguide resonators}
\author{Jeremy M. Sage}
\email{jsage@ll.mit.edu}
\affiliation{Lincoln Laboratory, Massachusetts Institute of Technology, Lexington, MA, 02420}
\author{Vladimir Bolkhovsky}
\affiliation{Lincoln Laboratory, Massachusetts Institute of Technology, Lexington, MA, 02420}
\author{William D. Oliver}
\affiliation{Lincoln Laboratory, Massachusetts Institute of Technology, Lexington, MA, 02420}
\author{Benjamin Turek}
\affiliation{Lincoln Laboratory, Massachusetts Institute of Technology, Lexington, MA, 02420}
\author{Paul B. Welander}
\affiliation{Lincoln Laboratory, Massachusetts Institute of Technology, Lexington, MA, 02420}

\date{\today}

\begin{abstract}
Superconducting coplanar waveguide (SCPW) resonators have a wide range of applications due to the combination of their planar geometry and high quality factors relative to normal metals.  However, their performance is sensitive to both the details of their geometry and the materials and processes that are used in their fabrication.  In this paper, we study the dependence of SCPW resonator performance on materials and geometry as a function of temperature and excitation power.  We measure quality factors greater than $2\times10^6$ at high excitation power and $6\times10^5$ at a power comparable to that generated by a single microwave photon circulating in the resonator.  We examine the limits to the high excitation power performance of the resonators and find it to be consistent with a model of radiation loss.  We further observe that while in all cases the quality factors are degraded as the temperature and power are reduced due to dielectric loss, the size of this effect is dependent on resonator materials and geometry.  Finally, we demonstrate that the dielectric loss can be controlled in principle using a separate excitation near the resonance frequencies of the resonator.
\end{abstract}

\maketitle

\section{Introduction}
There has been growing interest in fabricating high-Q superconducting coplanar waveguide (SCPW) resonators because of the many applications for which they may be used, such as narrow bandwidth microwave filters,\cite{Simon:filterreview:IEEE04} energy-resolving photon detectors,\cite{Day:KID:Nat03} and sub-quantum limited parametric amplifiers.\cite{Lehnert:Paraamp:APL07, Devoret:Paraamp:Nat10, Tholen:ParaampWL:PScripta09}  In the field of quantum information science, high-Q SCPW resonators play an extremely important role in superconducting qubit design\cite{Schrier:Transmon:PRB08}, inter-qubit coupling,\cite{Majer:QCoupler:Nat07} quantum information storage,\cite{Koch:Parking:PRL06} and quantum-state dispersive readout.\cite{Walraff:DispRd:PRL05, Metcalf:CavJBA:PRB07}  In addition, because superconducting qubits are fabricated using the same materials and processes as resonators, the study of the loss mechanisms limiting the Q in these resonators may prove to be a useful and relatively simple tool for understanding the fabrication-dependent limits to qubit coherence times.\cite{Martinis:TLSDec:PRL05, o'connellMart:SPLoss:APL08, wangMart:ResLoss:APL09, Lindstrom:ResLoss:PRB09, barends:NbTiN1:APL08, barends:NbTiNgrooves:APL10}

For quantum information science applications at microwave frequencies (e.g., solid state semiconductor and superconducting qubits), the resonators must be operated in a regime of low temperature ($\sim$10 mK) and low excitation power (single photon).  However, it has been found that in this regime the resonator quality factors are degraded as compared with their high temperature and/or high power values and that the dominant loss mechanism can be attributed to the presence of quantum two-level systems (TLS) in the resonator dielectrics.\cite{o'connellMart:SPLoss:APL08, wangMart:ResLoss:APL09, Lindstrom:ResLoss:PRB09, barends:NbTiN1:APL08, barends:NbTiNgrooves:APL10, gao:freqvsT:APL08}  These TLS may reside in the bulk dielectric substrate, at the metal-dielectric interface or in a thin layer on the metal and/or dielectric surfaces exposed to air.

In this paper, we study the excitation power dependence of the Q values  of SCPW resonators fabricated from Nb, Al, Re, and TiN metals deposited on sapphire, Si, and SiO$_2$.  We show that this dependence is consistent with the TLS model\cite{Phillips:TLSTheory:RPP87} and is affected by both the materials used and the geometry of the resonators.  We further show that we can measure the temperature dependence of the resonator's resonance frequency at relatively high temperature and excitation power and, using this TLS model, extract the low power and low temperature loss.  We examine the higher power performance of the resonators and find evidence for the Q being limited by geometry-dependent radiation loss.  Finally, we study the effect of controlling the dielectric loss by using a separate excitation near the resonator resonance frequencies to saturate, or pump, the TLS that interact with a much weaker resonant probe excitation of the resonator.

\section{Device Design and Measurement}
\begin{figure}

\includegraphics[width=.5\columnwidth]{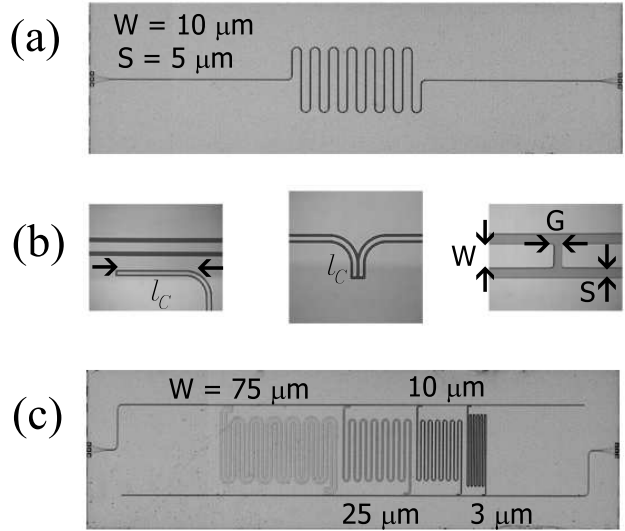}

\caption{\label{fig:ResPics} Resonator geometry and design.  (a) A single resonator device is shown that has W = 10 $\mu$m and S = 5 $\mu$m.  The resonator is meandered to keep the chip size small enough to avoid package resonances near the first few harmonics of the resonator.  There is a gap of size G = 4 $\mu$m in the straight section of this design that cannot be seen at this resolution, which couples the feedline to the resonator. (b) Zoomed in views of the resonator-to-feedline coupling are shown.  The left and center pictures show an L-coupler of length $l_C$ of the type used for the multiple resonator devices and single resonator devices, respectively.  The right picture shows a gap coupler of size G that has a width equal to that of the resonator and feedline center trace, W.  The spacing S between ground planes and center conductor is also shown.  (c) A multiple resonator device is shown with four resonators having different lengths for frequency multiplexing. The length difference is chosen so that adjacent resonators have frequencies which differ by $\sim$ 100 MHz.  The particular device shown has resonators of different widths with W = 3, 10, 25 and 75 $\mu$m.  The full chip ((a) and (c)) is $\sim$ 5 $\times$ 15 mm.}

\end{figure}

The devices we measure are half-wavelength ($\lambda/2$) SCPW resonators which have the general geometries shown in Fig. \ref{fig:ResPics}.  The ratio of the center conductor width W to the separation between the center conductor and ground planes S is designed so that the resonator transmission lines have a $\sim$50 $\Omega$ impedance, except in the investigation of radiation loss where we vary S independently of W.  The resonator lengths are chosen so that the fundamental resonance frequencies are 2-3 GHz, and the resonators have a meandered geometry so that their fundamental frequency and first few higher harmonics are below any package resonances.  The devices are either single resonator structures or multiple resonator structures.  The single resonator devices (Fig. \ref{fig:ResPics}(a)) are capacitively coupled to the input and output feedlines using either a gap of length G and width W or an L-coupler of length $l_C$ (see Fig. \ref{fig:ResPics}(b)).  The multiple resonator structures (Fig. \ref{fig:ResPics}(c)) have resonators of slightly different lengths so that they can be frequency multiplexed allowing for individual excitation of each resonator using common input and output feedlines coupled to each resonator with an L-coupler.  The device chips are 5 $\times$ 15 mm and are mounted with silver paint in a closed Au-plated OFHC copper package.  Connections to the input and output of the devices on chip are made through K-connectors on the package that have their center pin soldered to one end of the center conductor of a section of coplanar waveguide (CPW) made from Ti/Au deposited on alumina; this center conductor tapers down in width as it approaches the device chip, maintaining a 50 $\Omega$ impedance.  The other end of the CPW is wirebonded to a Ti/Au pad on the center conductor of the SCPW feedline on the resonator chip.  The chip and the Ti/Au on alumina CPW sections are grounded by making several hundred wirebond connections between the package and the Ti/Au and superconducting metal ground planes.

All low power measurements were performed in a dilution refrigerator at a base temperature of $<$ 10 mK.  The excitation signal is provided by the output port of a vector network analyzer.  The signal passes through $\sim$ 30 dB of cold attenuation on the input side to the device.  The output of the device goes through one of six inputs of a SP6T switch mounted at the mixing chamber.  The output of the switch is then sent through $\sim$ 3 dB of attenuation (used to thermalize the lines) to a cryogenic amplifier (mounted at 4 K) which has a bandwidth of 1-10 GHz and gain of $\sim$ 30 dB.  The noise temperature of the amplifier was measured at 4.2 K to be $\sim$ 25 K from 2-8 GHz.  The output signal then goes through $\sim$ 60 dB of room temperature amplification before entering the input port of the network analyzer.  The $S_{21}$ magnitude versus frequency is measured and saved for later analysis.  The measurements of resonators in which the temperature was varied were performed in a $^3$He refrigerator with a base temperature of 260 mK and having similar input and output attenuation and amplification as the dilution refrigerator.

The coupling of feedlines to all resonators is designed to be weak so that the internal quality factor $Q_I$ is approximately equal to the loaded quality factor $Q_L$.  This is done at the expense of signal to noise so that the extracted value of $Q_I$ is relatively insensitive to the calibration of the loss in the feedlines.  The weak coupling has the added benefit of reflecting the bulk of the 25 K noise coming from the cryogenic amplifier input so that there is at most only a few photons of noise on average in the cavity when we probe at low excitation power even without the use of an isolator, circulator, or significant attenuation on the output lines between the device and amplifier.  The calculation of the amount of noise entering the resonator due to the weak coupling is treated in the appendix.  The value of $Q_I$ is extracted from the measured value $Q_L$ through the relation
\begin{equation}
Q_I = \frac{Q_L}{1-10^{-\frac{IL}{20}}},
\label{eq:QIdef}
\end{equation}
where $IL$ is the insertion loss measured (in dB) once all coaxial lines leading to and from the resonator have been calibrated.  This calibration is done by measuring with the network analyzer the transmission through a section of superconducting coplanar waveguide that has no coupling capacitor gaps in the center trace and that is packaged in the same manner as the resonators.  The input of this calibration device is connected to a different but nominally identical refrigerator coax line while its output is connected through the cryogenic switch to the same output coax line as the resonators.  This allows us to calibrate the lines and the network analyzer each cool-down.  By changing which input line is used for the calibration device on multiple cool-downs and comparing the measured loss through the calibration device, we find a variability of $<$ 2 dB indicating our calibration of the $IL$ is accurate to within this degree of uncertainty.  In our resonator designs, the $IL$ at high excitation power is always, by design, $\geq$ 20 dB (at lowest probe power, the $IL$ increases due to added internal loss and is always $\geq$ 30 dB); differentiating Eq. (\ref{eq:QIdef}) with respect to $IL$ gives an estimated error in the value of $Q_I$ of at most $3 \%$ given our typical uncertainty of $\sim$ 2 dB in our measurement of $IL$.  In addition to a calibration of the $IL$, we also measure the transmission through the calibration device that has its input and output connected to different but nominally identical coax lines and divide the total measured round-trip insertion loss by a factor of 2.  This value is then subtracted from the value output by the network analyzer and represents the power incident on the input capacitor of the resonator, $P_{inc}$; we reasonably assume the level of uncertainty in $P_{inc}$ to be that found in the calibration of $IL$.

\section{Device Fabrication}
Our devices are fabricated using either Nb, Al, Re, or TiN films deposited on sapphire or on high-$\rho$ Si that has a 500 nm layer of ``wet" thermal SiO$_2$ (grown at elevated temperatures in a diffusion furnace), ``dry" SiO$_2$ (grown in dry oxygen), or that has its natural oxide removed and is subsequently hydrogen passivated just prior to metal deposition to minimize the formation of an oxide between the Si and the metal film.  The metal films were grown to be either poly-crystalline or epitaxial for Nb and Al, only epitaxial for Re, and only poly-crystalline for TiN.  For poly-crystalline Nb and Al deposited on Si, the metals were sputtered at pressures between 1.5 and 2 mTorr at room temperature with thicknesses of 200 nm and 300 nm, respectively.  Patterning was performed with I-line photoresist SPR511 followed by reactive ion etch (RIE) in fluorine-based chemistry for Nb and chlorine based chemistry for Al.  For poly-crystalline Nb and Al deposited on sapphire, the wafers were cleaned prior to metal deposition using SC1/Piranha chemistry.  The Nb and Al were sputtered at 10 mTorr with thickness of 200 nm and 6 mTorr with thickness of 300 nm, respectively.  Patterning was done with EZ1450 photoresist followed by RIE in fluorine-based chemistry for Nb and wet etch for Al.  The difference in processing between the two types of substrates arises because of the capabilities of different tools involved in handling 150 mm wafers (Si) and 50 mm wafers (sapphire).  The TiN films were deposited by reactive sputtering of Ti in N$_2$/Ar using 80$\%$ N$_2$ partial pressure and were patterned using I-line SPR511 photoresist and RIE in chlorine-based chemistry.  All epitaxial films were grown by molecular beam epitaxy on sapphire following a SC1/Piranha clean of the wafers.  Epitaxial Nb (110) was grown on A-plane sapphire with a thickness of 200 nm followed by a 1 nm Al$_2$O$_3$ (001) layer at 800 $^\circ$C.  Epitaxial Al (111) and Re (001) were grown on C-plane sapphire with thicknesses of 200 nm (at a temperature of 20 $^\circ$C) and 150 nm (at a temperature of 850 $^\circ$C), respectively.  The epitaxial Nb and Al were patterned in the same manner as the poly-crystalline Nb and Al on sapphire.  The Re was patterned using I-line photoresist SPR511 and RIE in fluorine-based chemistry.

\section{Results of Dielectric Loss Measurements: Low Power Regime}
\begin{figure*}

\includegraphics[width=\columnwidth]{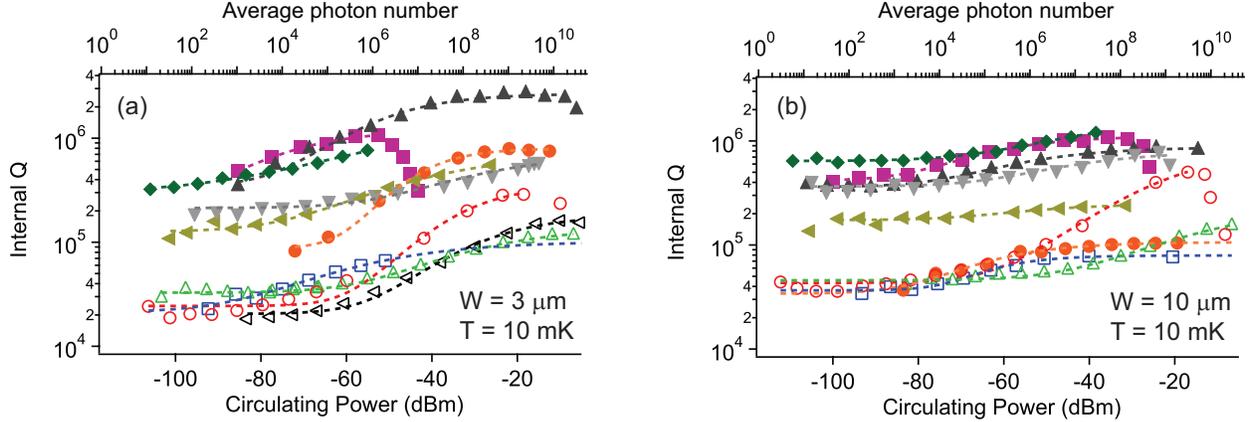}

\caption{\label{fig:QvsP} (Color online) Internal Q versus circulating power and equivalent average photon number for resonators of different materials for (a) W = 3 $\mu$m and (b) W = 10 $\mu$m.  The symbols indicate the different materials: $(\circ)$ poly-Nb/``wet" SiO$_2$/Si; $(\square)$ poly-Nb/Si; $(\vartriangle)$ poly-Nb/Sapphire; $(\triangleleft)$ epi-Nb/Sapphire; $(\bullet)$ poly-Al/``dry" SiO$_2$/Si; $(\blacktriangle)$ poly-Al/Sapphire; $(\blacksquare)$poly-Al/Si; $(\blacktriangleleft)$ epi-Al/Sapphire; $(\blacktriangledown)$ epi-Re/Sapphire; $(\blacklozenge)$ TiN/Si.  The dashed lines are fits to Eq. (\ref{eq:QItotal}).}

\end{figure*}

In Fig. \ref{fig:QvsP} we show a summary of the results of resonator performance as a function of circulating power and equivalent average photon number, $N_{ph}$.  This photon number is calculated from the circulating power through $N_{ph}=\frac{P_{circ}}{hf_0^2}$, where $f_0$ is the frequency of radiation probing the resonator.  In each graph is plotted $Q_I$ vs circulating power in the resonator $P_{circ}$ for the various combinations of metal and substrate and for a choice of W = 3 $\mu$m and S=2 $\mu$m (Fig. \ref{fig:QvsP}(a)) and W = 10 $\mu$m and S=5 $\mu$m (Fig.\ref{fig:QvsP}(b)).  The data for each point are obtained by fitting the magnitude of the measured $S_{21}$ versus frequency trace to a complex Lorentzian given by
\begin{equation}
S_{21}(f) = 10log_{10}\left|\frac{10^{-\frac{IL}{10}}}{1+2jQ_L^2\frac{f-f_0}{f_0}}+Ae^{j\phi}\right|,
\label{eq:S21def}
\end{equation}
where $f$ is the excitation frequency and $f_0$, $Q_L$, $IL$, $A$, and $\phi$ are the fit parameters.  The second term in Eq. (\ref{eq:S21def}) is a complex background that must be included to account for radiation coupling directly from input to output that bypasses the resonator.  The parameters $f_0$, $Q_L$, and $IL$ are the extracted values for the resonance frequency, loaded Q value, and insertion loss, respectively.  Once the fit is performed, the value of $Q_I$ is calculated from Eq. (\ref{eq:QIdef}) and $P_{circ}$ is calculated from the formula
\begin{equation}
P_{circ}=P_{inc}Q_L\frac{10^{-\frac{IL}{20}}}{n\pi},
\label{eq:Pcircdef}
\end{equation}
where $n$ is the $n^{th}$ harmonic being probed.  It is worth noting that $Q_L$ and $IL$ are not independent parameters; one of these parameters may be constrained if it is assumed that the external quality factor $Q_{ext}$ is determined solely by geometry and is independent of power and temperature.  This can be seen by using $\frac{1}{Q_{L}}=\frac{1}{Q_{I}}+\frac{1}{Q_{ext}}$ and Eq. (\ref{eq:QIdef}) to see that $Q_{L}=Q_{ext}10^{-\frac{IL}{20}}$.  One can therefore in principle fit a trace taken at high excitation power (where the signal to noise is high and the error is minimized) and use the extracted fit values to constrain the fits of data taken at low power in order to reduce fit errors.  We verified that, for a given resonator, the extracted values of $Q_{ext}$ were constant to within the uncertainty of our fit values over the entire range of $P_{inc}$; this indicates that indeed $Q_{ext}$ depends only on geometry.  An examination of this technique of constraining one fit parameter showed that fit errors could be reduced by roughly a factor of two.

For all materials and geometries, $Q_{I}$ decreases as $P_{circ}$ decreases which is consistent with TLS becoming unsaturated at low power (Fig. \ref{fig:QvsP}).  For many of the devices tested, we are able to observe $Q_I$ level off at low power which we associate with complete desaturation of TLS (unsaturated $Q_I$).  The positions of the left axes in Figs. \ref{fig:QvsP}(a) and (b) roughly indicate the value of circulating power corresponding to an average of one photon in the resonator; this value is given by $P^{1ph}_{circ}=nhf_{0}^2$ and is $\sim$ -116 dBm for the fundamental harmonic at 2 GHz.  For a particular resonator width, the trends of $Q_I$ at low power for the various materials appears to fall into a bimodal distribution with the Nb resonators having a lower average unsaturated $Q_I$ than the Al, Re, and TiN resonators.  This unsaturated $Q_I$ does not appear to depend strongly on the substrate; the strongest correlation to increased loss is the type of metal with which the resonator is fabricated.  This suggests that the dominant source of loss at low power comes from TLS in a dielectric formed on the metal, either at the metal/air interface or the metal/bulk dielectric interface.  It is known that Nb forms a more substantial oxide when exposed to room temperature air\cite{Halbritter:NbOxide:87} than do Al,\cite{Benndorf:AlOxide:77} Re,\cite{Ducros:ReOxide:80} and TiN,\cite{Glaser:TiNOxide:07} which all have a thin self-limiting oxide; this therefore points to an oxide at the metal/air interface as the most likely source of TLS in these resonators.  We have performed X-ray photelectron spectroscopy (XPS) on our post-patterned films and we have explicitly verified the presence of a surface oxide of Nb$_2$O$_5$, Al$_2$O$_3$, ReO$_2$, and TiO$_2$ on the Nb, Al, Re, and TiN films, respectively.  This technique did not allow us to determine the relative thicknesses of the oxides but it provides further evidence that the TLS loss arises from a surface oxide in all films.  The only data that are in contrast to this claim are those from the Al/SiO$_2$/Si resonators which have a loss comparable to the Nb resonators; however, it is known that SiO$_2$ is a large source of TLS\cite{Martinis:TLSDec:PRL05} and for these resonators the SiO$_2$ loss likely dominates the loss from the Al surface oxide.  It has been suggested that OH groups might be a source of TLS; \cite{Hunklinger:OH:76} indeed, this was our main motivation for growing a hydrogen-free ``dry" oxide.  However, since we observe comparable loss in the ``dry" and ``wet" oxide, it appears  that these OH groups are not a dominant TLS source.

The highest unsaturated $Q_I$ that we have measured is in the $W=10$ $\mu$m poly-crystalline TiN on Si with a value $Q_I=650,000$, comparable to the value reported in Ref. \onlinecite{Pappas:TiN:10}.  The TiN may exhibit a lower loss than Re and Al due to a thinner (or lower loss) metal oxide.\cite{Lkinnote}  In addition to suggesting the promise of using nitride superconductors in low-loss quantum circuits, the improved performance of TiN suggests that the Re and Al resonators are not being limited by the substrate; rather, their performance must be largely limited by the metal since the TiN is deposited on the same substrate.  As shown in Fig. \ref{fig:QvsP}(b), the poly-crystalline Al on sapphire and high-$\rho$ Si have comparable unsaturated Q values of 400,000. Since the two Al metals were deposited in different systems and etched using different methods, this further suggests that the loss limiting the Q is relatively insensitive to the processing conditions.  Furthermore, the fact that the Q is comparable for Al deposited on different substrates suggests once again that the metal/air interface is the dominant source of TLS.

In the usual model of TLS, the power-dependent loss is given by\cite{Phillips:TLSTheory:RPP87}
\begin{equation}
\frac{1}{Q_d(P)}=\frac{V_f}{Q_d^0}\frac{\mathrm{tanh}\left[\frac{hf_0}{2k_BT}\right]}{\sqrt{1+\frac{\mathrm{P}}{\mathrm{P_C}}}},
\label{eq:QvsPexact}
\end{equation}
where $\frac{1}{Q_d^0}$ is the unsaturated loss tangent of the dielectric due to TLS, $V_f$ is the fraction of the total volume the electromagnetic field occupies that contains TLS (this assumes the density of TLS is constant in the region $V_f$), $T$ is the temperature, and $P_C$ is a critical power that is determined by the transition strength of the TLS.  The power $P$ is not simply $P_{circ}$ since even if one assumes that the density of TLS is uniform over the volume $V_f$, the microwave field due to $P_{circ}$ is not uniform.  Rather, it varies both sinusoidally down the length of the resonator due to the standing wave that is present and varies in the directions orthogonal to the resonator trace due to the non-uniform field distribution generated in a CPW structure.  As a result, Eq. (\ref{eq:QvsPexact}) cannot be used directly to fit the data of Fig. \ref{fig:QvsP}.  By assuming standing wave potentials on the metal surfaces, one can numerically calculate the field distribution everywhere in space and show that the data can be fit by integrating Eq. (\ref{eq:QvsPexact}) weighted by the ratio of the local-to-total power at each point in $V_f$.\cite{gao:lossmodel:apl08, wangMart:ResLoss:APL09}  However, it has been shown the data can be consistently and more simply fit to a modified version of Eq. (\ref{eq:QvsPexact}),\cite{wangMart:ResLoss:APL09}

\begin{equation}
\frac{1}{Q_d(P_{circ})}=\frac{V_f}{Q_d^0}\frac{\mathrm{tanh}\left[\frac{hf_0}{2k_BT}\right]}{\sqrt{1+\left(\frac{P_{circ}}{\mathrm{P_C'}}\right)^{\beta}}},
\label{eq:QvsPapprox}
\end{equation}
where $\beta$ is a fit parameter and $P_C'$ is a critical power that now depends on the geometry of the resonator in addition to the microscopic properties of the TLS.  We fit the data in Fig. \ref{fig:QvsP} to

\begin{equation}
Q_{I}(P_{circ})=\frac{Q_d(P_{circ})Q^0}{Q_d(P_{circ})+Q^0},
\label{eq:QItotal}
\end{equation}
where $\frac{1}{Q^0}$ is a power independent loss that ultimately limits the resonator $Q_I$ when the TLS are saturated and $Q_d$$(P_C)$ is given by Eq. (\ref{eq:QvsPapprox}).  The fits are indicated by the dashed lines in Fig. \ref{fig:QvsP} and agree well with the data.  From the fits we extract the values of $\frac{V_f}{Q_d^0}$, $Q^0$, $P_C'$, and $\beta$.  Of particular interest is the parameter $\frac{V_f}{Q_d^0}$, since it directly determines the limit to the resonator $Q_I$ due to unsaturated TLS and is independent of the other loss mechanisms.  This parameter allows us to compare the TLS effects coming from different materials directly and the values obtained from the fits are shown in Table \ref{tab:tand} as the effective loss tangent tan$\delta$$_{\mathrm{eff}}$ $\equiv$ $\frac{V_f}{Q_d^0}$.  As already seen qualitatively in Fig. \ref{fig:QvsP}, the values of tan$\delta$$_{\mathrm{eff}}$ are bimodal with the resonators made from Al, Re, and TiN having roughly an order of magnitude lower value than those made with Nb.  It is also interesting to note that here the values for Al and Re are comparable, in apparent contrast with what was observed in Ref. \onlinecite{wangMart:ResLoss:APL09} where Re tended to have lower loss.  Furthermore, as expected from the higher measured $Q_I$, the value of tan$\delta$$_{\mathrm{eff}}$ = $9.6 \times 10^{-7}$ for the W = 10 $\mu$m TiN resonator is the lowest of all devices measured.

\begin{table}
\begin{tabular}{c|c|c|c}
  \hline
  \hline
  Metal & Dielectric & W ($\mu$m) & tan$\delta$$_{\mathrm{eff}}$ \\
  \hline
  Nb (poly) & ``wet" SiO$_2$/Si & 10 & 2.4e-5 \\
  Nb (poly) & Si & 10 & 1.5e-5 \\
  Nb (poly) & Sapphire & 10 & 1.8e-5 \\
  Al (poly) & ``dry" SiO$_2$/Si & 10 & 2.0e-5 \\
  Al (poly) & Si & 10 & 1.5e-6 \\
  Al (poly) & Sapphire & 10 & 1.6e-6 \\
  Al (epi) & Sapphire & 10 & 1.8e-6 \\
  Re (epi) & Sapphire & 10 & 1.8e-6 \\
  TiN (poly) & Si & 10 & 9.6e-7 \\
  \hline
  Nb (poly) & ``wet" SiO$_2$/Si & 3 & 5.7e-5 \\
  Nb (poly) & Si & 3 & 3.7e-5 \\
  Nb (poly) & Sapphire & 3 & 2.3e-5 \\
  Al (poly) & ``dry" SiO$_2$/Si & 3 & NA \\
  Al (poly) & Si & 3 & NA \\
  Al (poly) & Sapphire & 3 & NA\\
  Al (epi) & Sapphire & 3 & 6.5e-6 \\
  Re (epi) & Sapphire & 3 & 3.3e-6 \\
  TiN (poly) & Si & 3 & 3.0e-6 \\
  \hline
  \hline
  \end{tabular}
  \caption{\label{tab:tand} Fit values for tan$\delta$$_{\mathrm{eff}}$.  The table shows the values obtained for tan$\delta$$_{\mathrm{eff}}$ $\equiv$ $\frac{V_f}{Q_d^0}$ from the fits shown in Fig. \ref{fig:QvsP}.  Note that for a given material, the value is always higher for W = 3 $\mu$m as compared with W = 10 $\mu$m.  Like the low power data of Fig. \ref{fig:QvsP}, the Nb values are typically an order of magnitude higher than those for Al, Re and TiN.  The lowest value is for TiN, while the Al and Re values are comparable.  The three instances where ``NA" appears correspond to cases where the errors in the fit values of tan$\delta$$_{\mathrm{eff}}$ are comparable to the values themselves and are thus unreliable; this is due to the fact that the data could not be taken at low enough power to begin to see the leveling off of the Q value.  This occurs because of lower signal to noise arising from too weak a coupling of the feedlines to the resonators.}
\end{table}

The value of $\frac{V_f}{Q_d^0}$ is geometry dependent through the factor $V_f$; that is, the resonator geometry determines the fraction of the field that interacts with a certain spatial distribution of TLS.   Therefore, studying the geometry dependence of the TLS effects can give insight into the spatial distributions of the TLS.  Indeed, it is evident from Fig. \ref{fig:QvsP} and Table \ref{tab:tand} that the low power $Q_I$ for a given material is consistently higher for the $W=10$ $\mu$m resonators than for the $W=3$ $\mu$m resonators.  This has been previously observed\cite{gao:freqvsT:APL08} and it has been shown to be consistent with a thin layer of TLS near the surface of the resonator (such as a metal oxide or a thin lossy dielectric).  This dependence on geometry arises due to the electromagnetic field being more localized near the surface of the device for resonators with narrower transverse dimension (smaller W and S) and thus having a higher percentage of the electromagnetic energy in the region where the TLS reside.

\begin{figure}

\includegraphics[width=.5\columnwidth]{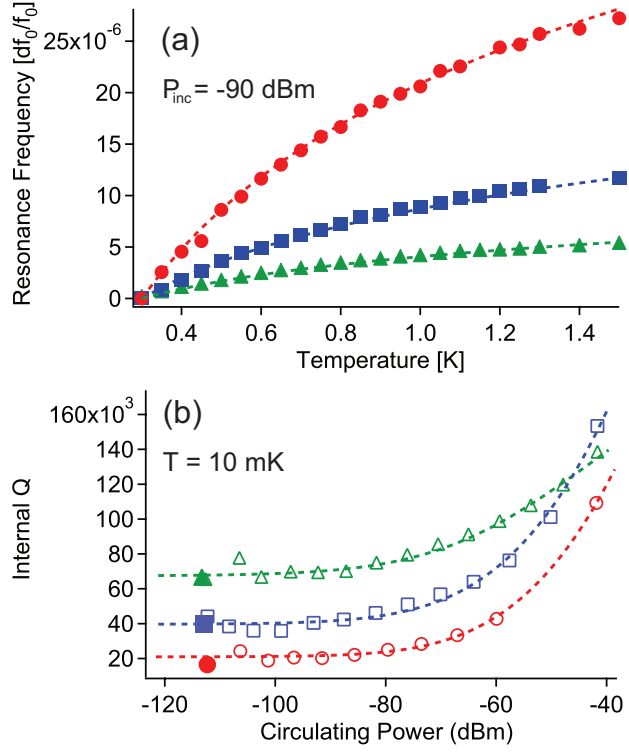}

\caption{\label{fig:WidthComp} (Color online) Comparison of resonators with varying width.  The data are taken using a Nb/SiO$_2$/Si multiplexed resonator device that has four resonators: W = 3, S = 2 $\mu$m; W = 10, S = 5 $\mu$m; W = 25, S = 15 $\mu$m; W = 75, S = 50 $\mu$m.  (The W = 75 $\mu$m resonator data are not shown due to its very low Q at all temperature and power levels).  The resonance frequencies are between 2-3 GHz. (a) The resonance center frequency versus temperature is shown for W = 3 $\mu$m $(\bullet)$, W = 10 $\mu$m $(\blacksquare)$, and W = 25 $\mu$m $(\blacktriangle)$.  The resonance frequency is expressed as a fractional change from the value at 300 mK.  The dotted lines are fits to the data using Eq. (\ref{eq:fvsT}).  It can be seen that the frequency dependence is stronger for narrower resonators.  (b) The internal Q versus circulating power is shown for W = 3 $\mu$m $(\circ)$, W = 10 $\mu$m $(\square)$, and W = 25 $\mu$m $(\vartriangle)$.  The dotted lines are fits to Eq. (\ref{eq:QItotal}).  It can be seen that the loss at low power is larger for narrower resonators.  Solid data points appear at the location corresponding to an average single-photon circulating power level; their values correspond to what we obtain by adding in parallel the tan$\delta$$_{\mathrm{eff}}^{-1}$  values extracted from the fits in (a) to the $Q^0$ values extracted from the fits in (b).  Doing this allows us to compare the low power internal Q value predicted from the frequency versus temperature data with the direct measurement of the low power internal Q.}

\end{figure}

To illustrate this in detail, we measured a Nb/SiO$_2$/Si multiplexed device with four resonators having different width ($W=3$, 10, 25, and 75 $\mu$m).  In Fig. \ref{fig:WidthComp}(a), we show the resonance frequency of the $W=3$, 10, and 25 $\mu$m resonators as a function of temperature ($Q_I$ of the 75 $\mu$m resonator was only $\sim$ 30,000 even at high power and this limitation prevented any measurement of effects from TLS).   In the TLS model, the resonance frequency of the resonator is expected to change with temperature as\cite{Phillips:TLSTheory:RPP87}

\begin{equation}
\frac{\Delta f}{f_0}=\frac{V_f}{\pi Q_d^0}Re\Psi \left[\frac{1}{2}+\frac{1}{2\pi i}\frac{hf_0}{k_BT}\right]-\ell n\left(\frac{hf_0}{k_BT}\right),
\label{eq:fvsT}
\end{equation}
where $\Psi(x)$ is the digamma function.  This dependence arises because the TLS become saturated for $k_BT$ $\gtrsim$ $hf_0$; this saturation leads to a dispersive frequency shift and is derived from the absorptive loss of Eq. (\ref{eq:QvsPexact}) through the Kramers-Kronig relations and by assuming that $P_{circ}$ is $\lesssim$ $P_C$.  The dashed lines in Fig. \ref{fig:WidthComp}(a) are the fits to Eq. (\ref{eq:fvsT}) from which we extract the values $\frac{V_f}{Q_d^0}$ = 5.7 $\times$ 10$^{-5}$, 2.4 $\times$ 10$^{-5}$, and 1.1 $\times$ 10$^{-5}$ for the W = 3, 10, and 25 $\mu$m resonators, respectively.  As expected, the change in frequency is larger for smaller W consistent with the TLS being distributed near the surface of the metal.  For this device, the TLS could be in a thin ($\sim$ few nm thick) oxide at one of the materials' interfaces, and/or uniformly distributed throughout the $\sim$ 200 nm thick SiO$_2$ between the Nb and Si.  Electromagnetic simulations \cite{Sonnet} show that the data are consistent with both distributions if one assumes a higher (lower) value for $Q_d^0$ for the thinner (thicker) oxide.   It is not surprising that both distributions can explain the data since even the smallest value of S (2 $\mu$m) is much greater than the ``thick'' oxide and one would not expect to be sensitive to resolving differences in TLS distributions on scales much less than S.  On the other hand, the thickness of the Si substrate (675 $\mu$m) is much greater than the values of S on the measured device and indeed the simulations show the data to be inconsistent with the TLS being distributed uniformly in the bulk Si substrate.

An interesting observation is that the value $\frac{V_f}{Q_d^0}$ extracted here at high temperature is the same value in Eq. (\ref{eq:QvsPapprox}) that determines the TLS loss at low power and low temperature.  Therefore, this analysis allows one to measure the temperature dependence of the resonance frequency at moderate power (since this dependence is power insensitive in such a regime) and relatively high temperatures ($>$ 300 mK) in order to determine the low temperature and low power $Q_I$ without working at dilution refrigerator temperatures or suffering the lower signal to noise that comes with probing at single photon-level powers.  The viability of this technique is shown in Fig. \ref{fig:WidthComp}(b), where we measure $Q_{I}$ as a function of power at low temperature and power.  The solid symbols show the values of $Q_I$ predicted from the fits in Fig. \ref{fig:WidthComp}(a), which are in excellent agreement with the measured values of $Q_I$ at low temperature and power.  Again, we see that the resonators with smaller W have higher loss.

\section{Results of Radiation Loss Measurements: High Power Regime}
\begin{figure}

\includegraphics[width=.8\columnwidth]{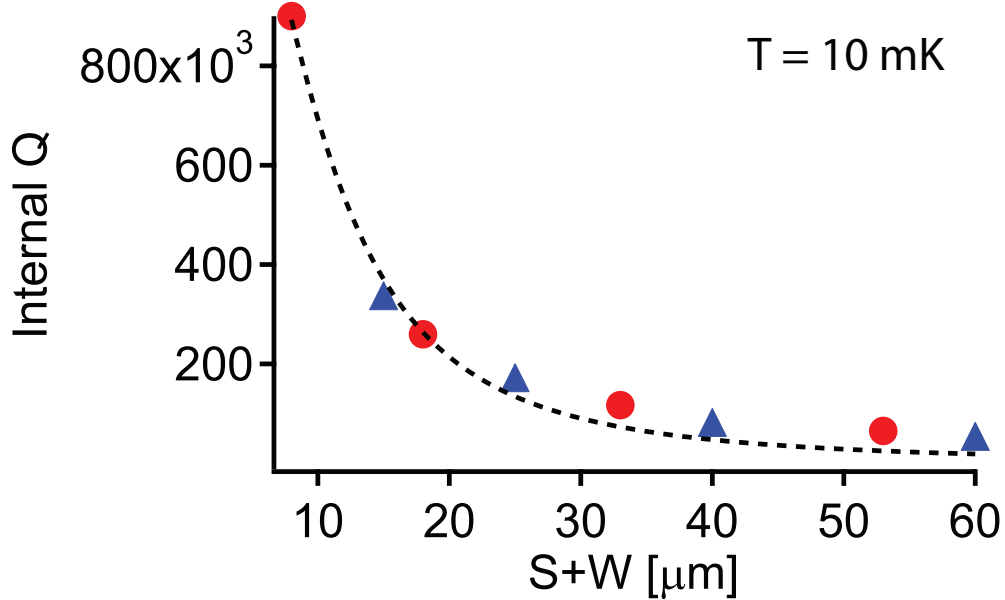}

\caption{\label{fig:Rad} (Color online) Internal Q versus the transverse resonator dimension, S+W.  The internal Q measured at high excitation power for eight resonators is plotted versus the value of S+W for each resonator.  The data are taken using two Re/sapphire multiplexed resonator devices having four resonators each.  Each device has resonators with S values of 5, 15, 30 and 50 $\mu$m and fixed width: W = 3 $\mu$m ($\bullet$) and  W = 10 $\mu$m $(\blacktriangle)$.  It can be seen that $Q_I$ decreases as S+W increases and that all the points appear to lie on a common curve.  Since the trend is observed for both devices separately, where W is fixed, the most likely explanation is a radiation loss that increases with S+W.  The dotted line is a fit to Eqs. (\ref{eq:Qrad1}) and (\ref{eq:Qrad2}) from which we extract a $\frac{1}{(S+W)^{2.3}}$ dependence of $Q_I$ arising from radiation loss Q$_{rad}$.}
\end{figure}

It is also interesting to examine the higher power performance of these resonators to determine the limit to $Q_I$ with the TLS saturated.  This limit may be imposed by loss due to residual quasiparticles in the superconductor, vortices, or radiation.  For the resonators measured in this work, the highest Q we obtain is for a W = 3 $\mu$m poly-crystalline Al on sapphire resonator with $Q_I$= 2.5 $\times 10^6$.\cite{Alsatcom}  We find that, in general, the highest Q values (at high power) are obtained in resonators with smaller widths W.  This result is qualitatively consistent with the limiting loss mechanism in resonators arising from vortex loss\cite{Plourde:Vortices:PRB09} in the film or radiation loss\cite{Mazin:thesis}, both known to increase as W increases.  These two effects can be separated by comparing resonators with fixed W and varying S.  Losses due to vortices should be insensitive to the value of S, while the radiation loss should increase as S increases\cite{Mazin:thesis}.  We perform this experiment with two separate frequency multiplexed resonator devices, both fabricated using epitaxial Re on sapphire.  The first device has four resonators with W = 3 $\mu$m and S = 5, 15, 30, and 50 $\mu$m, and the second device has four resonators with W = 10 $\mu$m and S = 5, 15, 30, and 50 $\mu$m.  We measure $Q_I$ as a function of circulating power for each resonator and record the maximum high power $Q_I$ value corresponding to the situation when the TLS are saturated.  We find that for both devices (i.e., both values of W), this $Q_I$ value decreases as S increases, which is consistent with loss due to radiation and not vortices.   Even more interestingly, when we plot the $Q_I$ values obtained from all eight resonators versus the value S+W, as shown in Fig. \ref{fig:Rad}, we see that the points appear to trend along a single curve.  We fit the data to the function

\begin{equation}
\frac{1}{Q_I}=\frac{1}{Q_0}+\frac{1}{Q_{rad}},
\label{eq:Qrad1}
\end{equation}
where $Q_0$ is a limiting loss mechanism that is independent of geometry and $Q_{rad}$ is a function of S+W, given by

\begin{equation}
Q_{rad}=\frac{\alpha}{(S+W)^{n_r}}.
\label{eq:Qrad2}
\end{equation}
From this fit, we extract for the fit parameters $\alpha$ = 2.8 $\times$ 10$^8$, $Q_0$= 1.5 $\times$ 10$^6$, and $n_r$ = 2.3.  The fit is indicated by the dashed line in Fig. \ref{fig:Rad} and it agrees well with the data.  The fit value $n_r$ = 2.3 is close to the $\frac{1}{(S+W)^2}$ dependence derived in Ref. \onlinecite{Mazin:thesis} where a straight (non-meandered) resonator geometry was assumed.  The radiation loss that occurs for higher values of S is unfortunate, since increasing S might otherwise have been a way to mitigate the effect of higher TLS loss in narrower resonators.  This, therefore, points out an important tradeoff that must be made in designing SCPW resonators: they must be narrow enough to avoid radiation loss, but wide enough to minimize loss from TLS at low power.

It is also worth noting that one difficulty with investigating the high-power $Q_I$, especially for the narrow $W=3$ and 10 $\mu$m resonators, is that one cannot completely saturate the TLS without locally heating the resonator and/or generating quasiparticles in the metal due to the large current densities flowing in the devices.  Heating of the device is verified by noting that as the input power is increased, the resonator center frequency initially shifts to a higher value. This direction of frequency shift is consistent with TLS being saturated with temperature according to Eq. (\ref{eq:fvsT}).  The quality factor increases, but the lineshape can become distorted due to the nonlinearity that results, making it impossible to fit the lineshape with a simple Lorentzian function and, thus, difficult to extract the  quality factor.  As the input power is increased, the resonance frequency begins to distort and shift to lower frequency, consistent with quasiparticle generation.  As a result, in some cases the loss from power induced quasiparticle generation cannot be completely deconvolved from the loss due to TLS.  This is not in principle a problem if one only wishes to determine the highest attainable $Q_I$ for given materials and geometry; for resonator design, this is an important figure of merit.  However, it demonstrates that in attempting to measure the high-power performance of SCPW, another design tradeoff is that the resonators should be narrow enough to avoid radiation loss while wide enough so that TLS can be saturated before encountering loss due to heating.

\begin{figure*}

\includegraphics[width=.8\columnwidth]{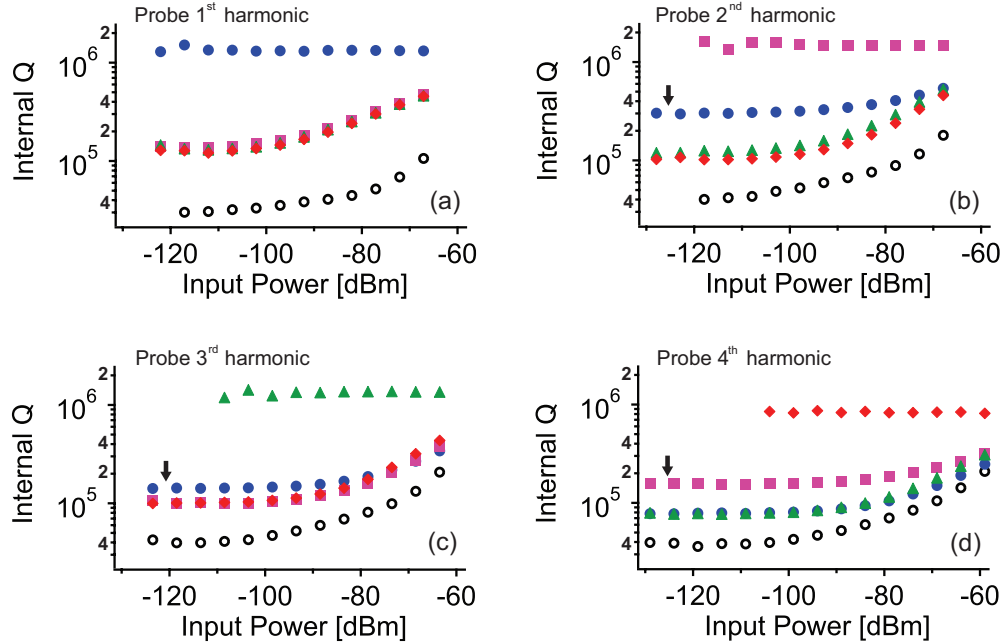}

\caption{\label{fig:PumpProbe} (Color online) Pump/probe measurements.  The first four harmonics of a W = 10, S = 5, G = 4 $\mu$m Nb/SiO$_2$/Si single resonator device are investigated.  The internal Q is plotted versus incident power of a ``probe" microwave tone for the (a) fundamental at 2.1381 GHz, (b) second harmonic at 4.2751 GHz, (c) third harmonic at 6.4116 GHz, and (d) fourth harmonic at 8.5485 GHz.  Each plot has five traces corresponding to the cases where either a second high power ``pump" tone is sent to the resonator with a frequency near the fundamental $(\bullet)$, second harmonic $(\blacksquare)$, third harmonic $(\blacktriangle)$, fourth harmonic $(\blacklozenge)$, or where there is no pump tone present $(\circ)$.  The detuning of the pump tone from its nearby resonance is chosen to be 20, 30, 40, and 50 kHz for the fundamental, second, third, and fourth harmonics, respectively.  The pump power value is chosen to be such that the high power quality factor saturates to its maximum value when the pump is near the same harmonic as the probe; here, the incident pump power P$_{inc}^{pump}$ is $\sim$ -20 dBm.  The data shows that in all cases, $Q_I$ is maximally enhanced by the presence of the pump whenever the pump frequency is near the same harmonic as the probe; this indicates that the TLS interact resonantly with the pump and probe.  We also observe enhancement of $Q_I$ when the pump is near a harmonic different from that of the probe. We claim that this enhancement is in general due to a heating effect, except in the cases where the probe frequency is near a multiple of the pump frequency.   We argue that the cause for the enhancement in these special cases is multi-photon driving of TLS.  The instances of n-photon driving are indicated by the black arrows in (b) and (d) for n = 2 and (c) for n = 3.}
\end{figure*}

\section{Two-tone Probing of Resonators}
Since it has already been shown that it is possible to saturate TLS at high power, it is interesting to explore the effect of saturating TLS with a separate microwave tone.  To do this, we combine a high-power tone coming from a variable frequency microwave signal generator (the ``pump'') with the tone coming from the network analyzer (the ``probe'') using a power combiner; the output of the combiner is then sent to the input of the resonator.   The resonator probed in this experiment is a Nb/SiO$_2$/Si device with W=10 $\mu$m and fundamental frequency of 2.14 GHz.  In order to couple the power of the pump into the resonator, the frequency of the pump is tuned to be a few linewidths from resonance with one of the first four harmonics of the resonator; the detuning is of order 10 kHz.  The probe frequency is then swept to measure the resonance of each harmonic as a function of probe power.  The pump power was increased until the TLS appeared to be saturated when probing the same harmonic as the pump.  Since the saturation of loss from TLS is a resonant effect\cite{Phillips:TLSTheory:RPP87} (i.e., only the TLS with energy splittings at or near the probe frequency will absorb and dissipate the power) it is expected that one should observe a reduction in loss whenever the pump is on resonance with TLS that have a splitting near the probe frequency.  As a result, this experiment investigates the possibility of whether the TLS are in fact multi-level systems (MLS) that may potentially be saturated by driving transitions between the ground state and higher lying excited states.  For instance, if such MLS were very weakly anharmonic, one might expect to see a reduction of loss when pumping at some multiple of the probe frequency.  In addition, this experiment is sensitive to the potential for multi-photon driving of TLS, which may be observed whenever the probe frequency is near an integer multiple of the pump frequency.  This pump/probe experiment is akin to hole-burning and is a kind of coarse saturation-absorption spectroscopy of TLS at four discreet frequency points.\cite{Demtroder:LaserSpec:96}

In Fig. \ref{fig:PumpProbe}(a)-(d), we show the results of the measurements, where we plot $Q_I$ vs the incident probe power $P_{inc}$ for the sixteen different combinations of pumping and probing near the first four harmonics.  Also shown is the effect of probing at each harmonic in the absence of the pump.  As can be seen from the data, for all four harmonics, $Q_I$ remains high even at low incident probe power whenever we pump near the same harmonic as we probe.  This demonstrates that TLS can indeed be saturated with a second tone at a frequency detuned from resonance.  In addition, we see an enhanced $Q_I$ at low probe power when we pump near harmonics differing from the ones being probed, as compared with the case where there is no pump.  In most cases, we believe this is due to heating of the resonator by the pump which acts as a ``broadband" saturation of TLS as predicted by the temperature dependence of Eq. (\ref{eq:QvsPexact}).  This is the most likely explanation since, for a given probe, there appears to be minimal difference between the curves for differing pump frequencies (see Fig. \ref{fig:PumpProbe}(a), for instance).  This similarity in the curves, indicating a lack of spectroscopic structure, is what one would expect from a broadband, non-resonant excitation of TLS.  Also, in all cases we observe that the resonance frequency shifts to a higher value in the presence of the pump which is consistent with an increased temperature.  However, in a few particular instances, we observe an enhancement of $Q_I$ above this background improvement due to heating.  This occurs in the cases whenever the probe frequency is near an integer multiple of the pump.  The cases where this enhancement occurs as we 1) probe at the second harmonic while pumping near the first and 2) probe at the fourth harmonic while pumping near the second correspond to two-photon driving of TLS.  The enhancement observed in the case where we probe at the third harmonic and pump near the first corresponds to three-photon driving.

In interpreting these results, one should in principle consider that they can also be explained by harmonic generation due to the high microwave power and nonlinearity of the superconducting material.  This explanation would assert that frequency doubled and tripled radiation drive the TLS rather than two-photon and three-photon driving.  However, we rule out this possibility for two reasons.  First, it is known that second harmonic generation is strongly suppressed in a superconductor in the presence of zero magnetic field.\cite{Gallitto:SHG:EPJB06}  Second, the resonator probed here is quite anharmonic; the center frequencies of higher harmonics are not integer multiples of the fundamental frequency, so much so that the anharmonicity ($>$ 1 MHz) is much greater than the linewidth of the resonances ($<$ 10 kHz).  As a result, any harmonic generation would produce tones that were very far off resonance with the resonator and their amplitude in the resonator would be strongly suppressed.

The multi-photon driving of TLS in the presence of the resonator anharmonicty allows us to put an approximate lower bound on the TLS linewidths.  In order to see a reduction of TLS loss, we must be saturating with the pump the particular TLS that are absorbing energy from the probe.  We can therefore infer that we must be effectively driving the TLS despite being detuned by the resonator anharmonicity.  Thus, we can say that the linewidths of the TLS are $\gtrsim$ 1 MHz, corresponding to a lifetime of $\tau_{TLS}$ $\lesssim$ 150 ns assuming $T_1$ is the dominant source of line broadening.  We note that we do not see any enhancement of $Q_I$ at low probe power when we pump at frequencies that are near multiples of the probe.  We therefore see no evidence of MLS indicating that either we are not effectively driving and/or detecting transitions to higher excited states, or that the sources of loss are indeed true TLS.

As a final point, this pump/probe experiment suggests that it may be possible to reduce the loss due to TLS, and the decoherence that results, in a general quantum circuit.  For instance, it has been shown that resonator quality factor is directly linked to single-photon Fock state lifetime in a resonator\cite{wangMart:ResLoss:APL09} and thus an enhancement of the quality factor may lead to longer coherence times, as suggested in Ref. \onlinecite{Dzero:PumpLoss:PRB10}.  Additionally, if a pump field can be made to overlap with a region containing TLS that are coupled to qubits, qubit coherence may be improved.  The demonstration of saturation of TLS with a pump frequency that differs from the resonance frequency of the circuit, either through off-resonant driving or through multi-photon excitation, could be practical in that the pump can be detuned from any quantum circuit element one might be concerned about spuriously exciting with the strong pump field.  Another idea suggested by this experiment is to use a pump of acoustic waves to saturate the TLS, as has been previously shown to work in principle.\cite{Doussineau:Cross:76, Hunklinger:Cross:77}  In this technique, one could potentially observe a decrease in loss in a quantum circuit with minimal spurious excitation of its elements since they should be decoupled to a good degree from phonons.  The idea of increasing coherence time through pumping of TLS is one that requires further investigation to assess its viability.

\section{Conclusion}
In conclusion, we have investigated the performance of SCPW resonators and find that it depends strongly on both materials and geometry.  We find that at low excitation power, the loss is enhanced due to the presence of TLS located on the surface of the superconducting metal and find that TiN has the lowest loss as compared with Nb, Al, and Re; Nb appears to suffer from a roughly order of magnitude higher loss than the other metals due to these surface TLS.   We find that the loss from TLS can be mitigated by increasing the width of the resonator and spreading the fields away from the surface where the TLS reside.  However, we find that by further increasing the width of the resonator one becomes limited by radiation loss, so that there is a tradeoff in design if high Q is desired at low power.  We demonstrate a technique in which we can extract the low power quality factor by measuring the temperature dependence of the resonance frequency at high power, thus providing an alternate method to investigate the low power performance.  Finally, we perform a pump/probe measurement in which we simultaneously excite the resonator with a strong pump tone and a weak probe tone.   We find that the TLS can be saturated by the pump tone when it is near in frequency to the probe or one of its subharmonics.  This technique can be used as a spectroscopic probe of TLS and potentially provides the means to control dielectric loss in resonators and other quantum circuit elements.

\appendix*
\section{}

There is in principle a concern that without sufficient attenuation or isolation between the cryogenic amplifier and the output to the device (or between the room temperature electronics and the input of the device), noise can reach the resonator and saturate the TLS.  If this were the case, one would not be able to accurately claim that the TLS loss measured with probe radiation corresponding to near single photon power level is the true low power loss.  However, we note that it is not the amount of noise impinging on the resonator from the outside that is of importance; rather it is the amount of noise that enters the resonator where the TLS reside that is of concern.  In the measurements discussed here we have only a small attenuation (3 dB) between our 25 K cryogenic amplifier and the device output, but we show below that our engineered weak coupling keeps the devices relatively immune from noise impinging from the amplifier.  To do this we assume that the noise is uncorrelated and incoherent. Under this assumption, there is no coherent power buildup in the resonator from the noise radiation.  We may then simply analyze the problem using a rate equation of energy (or photons) leaking into the resonator through the partially transmitting input mirror (or coupling gap capacitor) and leaking out through the input and output mirrors and through the internal loss channels of the resonator.  The rate equation is

\begin{equation}
\frac{d(E_{circ})}{dt} = P_{circ}-E_{circ}(\gamma_1+\gamma_2+\gamma_l),
\label{eq1}
\end{equation}
where $E_{circ}$ and $P_{circ}$ are the energy and power inside the resonator, respectively and $\gamma_1$, $\gamma_2$, and $\gamma_l$ are the rates of energy leaking out (and in) through the input mirror, output mirror, and through all internal loss channels, respectively.  In the steady state, we set $\frac{d(E_{circ})}{dt}=0$ and find

\begin{equation}
E_{circ}=\frac{P_{circ}}{\gamma_{tot}},
\label{eq2}
\end{equation}
where $\gamma_{tot}=\gamma_1+\gamma_2+\gamma_l$.

We can relate the circulating power inside the resonator $P_{in}$ to the energy incident on the resonator from the outside, $E_{inc}$ as

\begin{equation}
P_{circ}=E_{inc}\gamma_1.
\label{eq3}
\end{equation}

Combining Eqs. \ref{eq2} and \ref{eq3}, we get

\begin{equation}
E_{circ}=E_{inc}\frac{\gamma_1}{\gamma_{tot}}.
\label{eq4}
\end{equation}

We can recast equation \ref{eq4} by noting that, trivially, $\frac{E_{circ}}{E_{inc}}=\frac{P_{circ}}{P_{inc}}$, where $P_{inc}$ is the power incident from the outside, to get

\begin{equation}
P_{circ}=P_{inc}\frac{\gamma_1}{\gamma_{tot}}.
\label{eq5}
\end{equation}

We now note the rate values ($\gamma_1$, $\gamma_2$, $\gamma_l$) can be related to the resonator Q simply as

\begin{equation}
\frac{\gamma_1+\gamma_2}{\gamma_{tot}}=\frac{Q_L}{Q_{ext}}.
\label{eq6}
\end{equation}

If we then assume that $\gamma_1=\gamma_2$ (reasonable since our coupling capacitors are identical) and we combine Eqs. \ref{eq5} and \ref{eq6} using the relationship between $Q_L$ and $Q_{ext}$, $Q_L=Q_{ext}10^{-\frac{IL}{20}}$, we get

\begin{equation}
P_{circ}=\frac{1}{2}P_{inc}10^{-\frac{IL}{20}}.
\label{eq7}
\end{equation}

This can be recast, as before, as

\begin{equation}
E_{circ}=\frac{1}{2}E_{inc}10^{-\frac{IL}{20}}.
\label{eq8}
\end{equation}

Now, we note that we care primarily about the noise photons that would interact with the same TLS as our coherent probe radiation; i.e., those photons with the same frequency as the resonance frequency of the resonator, $f_0$.  The energy $E_{circ}$ can be written in terms of photon number, $N_{noise}$, as

\begin{equation}
E_{circ}=N_{noise}hf_0,
\label{eq9}
\end{equation}

where h is Planck's constant.  Finally, combining Eqs. \ref{eq8} and \ref{eq9}, we get

\begin{equation}
N_{noise}=\frac{1}{2}\frac{E_{inc}10^{-\frac{IL}{20}}}{hf_0}.
\label{eq10}
\end{equation}

Examining Eq. \ref{eq10}, we see that the noise inside the resonator is reduced by the factor $\frac{1}{2}10^{-\frac{IL}{20}}$ from that which is incident upon it, which is just a mathematical statement that weakly coupling the resonator to the feedlines weakly couples the noise as well.  The value of $E_{inc}$ is obtained for noise coming from the cryogenic amplifier with noise temperature $T_n$ using $E_{inc}$ = $\frac{1}{2}k_BT_n$, where $k_B$ is Boltzman's constant and the factor of $\frac{1}{2}$ is due to the 3 dB of attenuation between our cryogenic amplifier and our device output.  Plugging in numbers to Eq. \ref{eq10} using $IL$ = 30 dB, $f_0$ = 2 GHz, and $T_n$ = 25 K, we find $N_{noise}\sim2$.   This is certainly of the same order as a single photon and is why we can accurately claim that the left axes of the plots in Fig. \ref{fig:QvsP} correspond roughly to that of a single photon of circulating power.  As a final note, we can see that our devices are also immune from room temperature noise incident on their input from the $\sim$300 K room temperature electronics.  Since there is 30 dB of attenuation between room temperature and our device input, the device sees an effective $T_n$=300 mK.  Plugging $T_n$=300 mK into Eq. \ref{eq10}, we see that $N_{noise}\ll1$.

\begin{acknowledgments}
We thank George Fitch, Peter Murphy, Richard Slattery, and Susan Cann for technical assistance and Daniel Oates, John Chiaverini, Jamie Kerman, and the Lincoln Laboratory QIS team for helpful discussions.  This work is sponsored by the Defense Advanced Research Projects Agency under United States Air Force Contract $\#$FA8721-05-C-0002.  Opinions, interpretations, recommendations and conclusions are those of the authors and are not necessarily endorsed by the United States Government.  Approved for public release, distribution unlimited
\end{acknowledgments}


\begin{thebibliography}{35}
\bibitem [{Lki()}]{Lkinnote}%
  \BibitemOpen
  \href@noop {} {}\bibinfo {note} {In principle it is possible that the
  enhanced Q for TiN arises due to the larger kinetic inductance of TiN as
  compared with Al and Re due to an increased electric field for a given power.
  However, from our measurements of the frequency shift of our TiN resontor
  with respect to that of our Al resonator of identical geometry and substrate
  (which we conservatively assume has a much lower kinetic inductance
  fraction), we extract the TiN kinetic inductance fraction of $\sim$ 50
  percent and find that the corresponding electric field intensity increases by
  3 dB. Since the TiN and Al curves in Fig.\ref{fig:QvsP}(b) are flat over more
  than 10 dB at lowest power, the difference in Q value between the TiN and Al
  resonators arising from this effect would be negligible.}\BibitemShut {Stop}%
\bibitem [{Als()}]{Alsatcom}%
  \BibitemOpen
  \href@noop {} {}\bibinfo {note} {We note that the high power Q saturates at a
  higher power for the Al on sapphire device as compared with the Al on silicon
  device. This could be due to a differing thermal link between each device and
  its package through its substrate. This difference could cause the number of
  quasiparticles generated by heat arising from the high power in the resonator
  to vary between the two devices and thus could lead to a different saturation
  value. However, we do not have a definitive explanation and further
  investigation would be required to understand this effect.}\BibitemShut
  {Stop}%
\end{thebibliography}

%

\end{document}